\documentclass{epl}
\def\hatA{{\bf\skew5\hat{\hbox{$A$}}}}
\def\hatB{{\bf\skew3\hat{\hbox{$B$}}}}
\def\hatD{{\bf\skew2\hat{\hbox{$D$}}}}
\def\ochi{\overline{\overline{\chi}}}
\def\ounit{\overline{\overline{1}}}
\newcommand{\TN}{\hbox{$T_{\rm N}^{}$}}

\title{Magnetic Raman scattering of the ordered
tetrahedral spin-$\frac{1}{2}$ clusters in
Cu$_2$Te$_2$O$_5$(Br$_{1-x}$Cl$_x)_2$ compounds}
\shorttitle{Raman scattering in
Cu$_2$Te$_2$O$_5$(Br$_{1-x}$Cl$_x)_2$}
\author{J. Jensen\inst{1} \and P. Lemmens\inst{2} \and C. Gros\inst{3}}
\institute{
  \inst{1} \O rsted Laboratory, Niels Bohr Institute, Universitetsparken 5,
  DK-2100 Copenhagen, Denmark.\\
  \inst{2} Max Planck Institute for Solid State Research,
D-70569 Stuttgart, Germany.\\
  \inst{3} Fakult\"at 7, Theoretische Physik,
         Universit\"at des Saarlands,
         D-66041 Saarbr\"ucken, Germany. }

\begin{document}

\maketitle

\begin{abstract}
Raman light-scattering experiments in the antiferromagnetic phase of the
Cu$_2$Te$_2$O$_5$(Br$_{1-x}$Cl$_x)_2$ compounds are analyzed in terms of a
dimerized spin model for the tetrahedral Cu-clusters. It is shown that the
longitudinal magnetic excitation in the pure Br system hybridizes with a
localized singlet excitation due to the presence of a Dzyaloshinskii-Moriya
anisotropy term. The drastic change of the magnetic scattering intensities
observed when a proportion of Br is replaced by Cl ions, is proposed to be
caused by a change of the magnetic order parameter. Instead of being
parallel/antiparallel with each other, the spins in the two pairs of
spin-$\frac{1}{2}$ order perpendicular to each other, when the composition
$x$ is larger than about 0.25.

\end{abstract}

The compounds Cu$_2$Te$_2$O$_5$Br$_2$ and Cu$_2$Te$_2$O$_5$Cl$_2$
are spin-$\frac{1}{2}$ Cu-systems with coupled spin tetrahedra.
The tetragonal $P\overline{4}$ crystal structure of the two
compounds and the susceptibility measurements are presented in
ref.\ \cite{b.John}. While the high-temperature susceptibility
approaches that of uncoupled spins, at low temperatures the
susceptibility is reduced and goes through a maximum at,
respectively, 30 and 23 K. The maximum indicates that the $2^4$
spin-$\frac{1}{2}$ states of the Cu-ions in the (distorted)
tetrahedral clusters are dimerized so to create a singlet ground
state separated from the excited states by a gap of about 40 K.
Both the susceptibility and the heat capacity measurements clearly
indicate a transition to an ordered phase at a temperature, 11.4 K
in the Br and 18.2 K in the Cl system \cite{b.PL1,b.PL2}, well
below that of the maximum. Some of the bulk properties and the
Raman-scattering results of the Br compound have been analyzed
successfully by Gros \etal~\cite{b.Gros} in terms of a dimerized
model for the four Cu spins of one tetrahedron, as determined by
the following Hamiltonian:
\begin{equation}
\label{e.1} {\cal H}_t=J_1^{}\,({\bf S}_1^{}+{\bf
S}_2^{})\cdot({\bf S}_3^{} +{\bf S}_4^{})+J_2^{}\,({\bf
S}_1^{}\cdot{\bf S}_2^{}+{\bf S}_3^{} \cdot {\bf S}_4^{})
\end{equation}
$J_1^{}$ and $J_2^{}$ are both positive. Defining the ratio
$r=J_2^{}/J_1^{}$ then the energy differences between the singlet
ground state $|s_1^{}\rangle$ and the excited states are:
$2(1-r)J_1^{}$ to a singlet $|s_2^{}\rangle$, $J_1^{}$ to a one
triplet state $|t_1^{}\rangle$, $(2-r)J_1^{}$ to two degenerate
triplets $|t_2^{}\rangle$ and $|t_3^{}\rangle$, and $3J_1^{}$ to a
quintuplet $|q\rangle$. This tetrahedral unit is coupled to the
neighboring ones, and the Heisenberg interaction assumed in ref.\
\cite{b.Gros} is
\begin{equation}
\label{e.2}
{\cal H}_{\rm MF}^{(Br)}
=-J_c^{}M_z^{}\big(S_1^z+S_2^z-S_3^z-S_4^z\big)+2J_c^{}M_z^2,
\qquad M_z^{}=\frac{1}{4}\big\langle
S_1^z+S_2^z-S_3^z-S_4^z\big\rangle
\end{equation}
within the mean-field (MF) approximation, i.e., the order
parameter is one where the spins on each of the pairs 1--2 and
3--4 are parallel, but antiparallel with respect to the other pair
of spins. This ordering takes full advantage of the
$J_1^{}$-interaction on the expense of the $J_2^{}$-interaction.
In the model of Gros \etal~\cite{b.Gros} for the Br compound
$r=0.66$ is smaller than 1, and the system is close to quantum
criticality with a coupling parameter $J_c^{}=0.85J_1^{}$, only
13\% larger than the critical value $J_c^{(qc)}=0.75J_1^{}$. The
model was investigated analytically neglecting the modifications
due to higher lying levels. The high-temperature susceptibility is
accounted for using $J_1^{}=47.7$ K, the N\'eel temperature
increases with an applied field, and the frequency of the
longitudinal magnetic excitation in the ordered phase is estimated
to be close to that observed by Raman spectroscopy. Here we shall
present a complete MF analysis, which includes the calculation of
the susceptibility and the bulk magnetization in the ordered phase
and the field dependence of \TN. It is performed numerically
accounting for all effects of the total level scheme. The
Raman-scattering cross section is derived within the RPA, and the
field-induced $|s_2^{}\rangle$ singlet excitation in the Raman
spectrum is determined to be due to an intrinsic
Dzyaloshinskii-Moriya (DM) anisotropy term in the Hamiltonian of
the Cu spins and not to a DM contribution to the Raman-scattering
operator as proposed in \cite{b.Gros}.

The bulk properties of Cu$_2$Te$_2$O$_5$(Br$_{1-x}$Cl$_x)_2$
change gradually with $x$. This suggests that the only
modification occurring is an increase of the effective $J_c^{}$ in
proportion to $x$, corresponding to the increase of \TN\ from 11.4
to 18.2 K. However, the Raman scattering experiments, and to some
extent also the magnetization measurements in the ordered phase,
indicate that the nature of the ordered phase is substantially
modified when $x$ becomes larger than about 0.25. After a thorough
analysis of alternatives to the $M_z^{}$ order parameter in
(\ref{e.2}), we conclude that the only one which is in accord with
the experiments is the following
\begin{equation}
\label{e.3} {\cal H}_{\rm MF}^{(Cl)}
=-J_c^{xy}M_{xy}^{}\big(S_1^x-S_2^x-S_3^y+S_4^y\big)+2J_c^{xy}M_{xy}^2,
\qquad M_{xy}^{}=\frac{1}{4}\big\langle
S_1^x-S_2^x-S_3^y+S_4^y\big\rangle
\end{equation}
The $M_{xy}^{}$ parameter describes an ordered state in which each
pair of spins, 1--2 or 3--4, is in an antiparallel configuration,
and the spins on the two different pairs are perpendicular to each
other. When each pair of spins is antiferromagnetically ordered,
the free energy contribution of the $J_1^{}$ interaction cancels
out to leading order. Thus the angle between the spins on the two
different pairs is left undetermined by the intra-tetrahedral
interactions in eq.~(\ref{e.1}), whereas the DM-anisotropy favors
the perpendicular configuration.

\section{The MF/RPA theory and the Raman-scattering cross section}
The ground state properties and the excitations of the tetrahedral
Cu-systems have been analyzed in terms of a general MF/RPA theory,
see for instance \cite{b.Jens}. The MF order parameter is
determined in a self-consistent fashion, from a numerical
diagonalization of the total MF Hamiltonian followed by a
calculation of the thermal expectation value of the order
parameter. The linear response at zero or non-zero frequency is
derived from the non-interacting susceptibility tensor
$\ochi\,{}^0(\omega)$, with the components $\chi_{AB}^0(\omega)$
determined in terms of the operators $\hatA$ and $\hatB$, see
section 3.3 in [5]. Introducing the MF-coupling constant $J_c^{}$
in (2) then the $3\times3$ susceptibility tensor $\ochi(\omega)$,
corresponding to $\hatA$ and $\hatB$ equal to the three components
of ${\bf S}_1^{}+{\bf S}_2^{}-{\bf S}_3^{}-{\bf S}_4^{}$, is
determined from the non-interacting one according to
$\ochi(\omega)=\ochi\,{}^0(\omega)\big[\,\ounit-(J_c^{}/4)
\ochi\,{}^{0}(\omega)\big]^{-1}$. Introducing a new, independent
operator $\hatD$, then the response $\chi_{D\!D}^{}(\omega)$ is
derived from the $4\times4$ susceptibility tensor defined in the
vector space consisting of the three Cartesian components plus
$\hatD$ as the fourth component. Hence $\chi_{D\!D}^{}(\omega)$ is
the 44 component of the interacting susceptibility tensor
determined from the non-interacting one as above, except that
$(J_c^{}/4)$ is replaced by a diagonal matrix with a zero in the
44 position and $(J_c^{}/4)$ in the remaining part of the
diagonal.

The Raman scattering cross section $R$ may according to Brenig and
Becker \cite{b.Bre} be written $\sum_{lm}a_{lm}^{}({\bf
e}_i^{}\cdot n_{lm}^{})({\bf e}_o^{}\cdot n_{lm}^{}){\bf
S}_l^{}\cdot{\bf S}_m^{}$, where ${\bf e}_i^{}$ and ${\bf e}_o^{}$
are the unit vectors of the incoming and outgoing electric field
and ${\bf n}_{lm}^{}$ are the unit vectors connecting
exchange-coupled sites. When the incoming and outgoing light are
polarized along the $c$ axis, then $R$ is proportional to
\begin{eqnarray}
\label{e.7}\sum_i\Big[\big({\bf S}_1^{i}&\!\!+\!\!&{\bf
S}_2^i\big)\!\cdot\! \big({\bf S}_3^i+{\bf
S}_4^i\big)+\alpha_0^{}\big({\bf S}_1^{i}+{\bf
S}_2^i\big)\!\cdot\! \big({\bf S}_3^{i-1}\!+{\bf
S}_4^{i-1}\big)+\,\alpha_0^{}\big({\bf S}_3^{i}+{\bf
S}_4^i\big)\!\cdot\!\big({\bf
S}_1^{i+1}\!+{\bf S}_2^{i+1}\big)+\cdots\Big]\nonumber\\
&{}&{\mathop\rightarrow^{\rm RPA}}  \sum_i\Big[\big({\bf S}_1^{i}+{\bf
S}_2^i\big)\cdot \big({\bf S}_3^{i}+{\bf S}_4^i\big) -2\alpha{\bf
M}\cdot\big({\bf S}_1^{i}+{\bf S}_2^i-{\bf S}_3^{i}-{\bf
S}_4^i\big)\Big]
\end{eqnarray}
This RPA expression includes all neighboring exchange-coupled pairs via the
introduction of an effective $\alpha$. Defining the operators $\hatD\equiv
\big({\bf S}_1+{\bf S}_2\big)\cdot \big({\bf S}_3+{\bf S}_4\big)$ and
$\hatA\equiv S_{1z}^{}+S_{2z}^{}-S_{3z}^{}-S_{4z}^{}$ then the scattering
intensities in the Raman experiments are determined by
\begin{equation}
\label{e.8} I(\omega)\propto
\frac{1}{1-e^{-\hbar\omega/k_BT}}\,{\rm
Im}\left\{\chi_{D\!D}^{}(\omega)+4\alpha^2M_z^2\chi_{AA}^{}(\omega)
-2\alpha M_z^{}\big[ \chi_{AD}^{}(\omega)+
\chi_{DA}^{}(\omega)\big]\right\}
\end{equation}
If $\alpha=J_c^{}/ 2J_1^{}$ then $R_{\rm RPA}^{}$ commutes with
the MF Hamiltonian. This corresponds to the case considered by
Brenig and Becker \cite{b.Bre} of $\beta=b$. The Raman-scattering
intensity vanishes identically in this case, independent of the
value of $M_z^{}$. The value of $\alpha$ is unknown, and in the
following we simply assume that $\chi_{D\!D}^{}(\omega)$ is the
dominating term, i.e., that $\alpha\approx0$.

\section{Analysis of the experiments}

\begin{figure}[t]
\twofigures[scale=0.42]{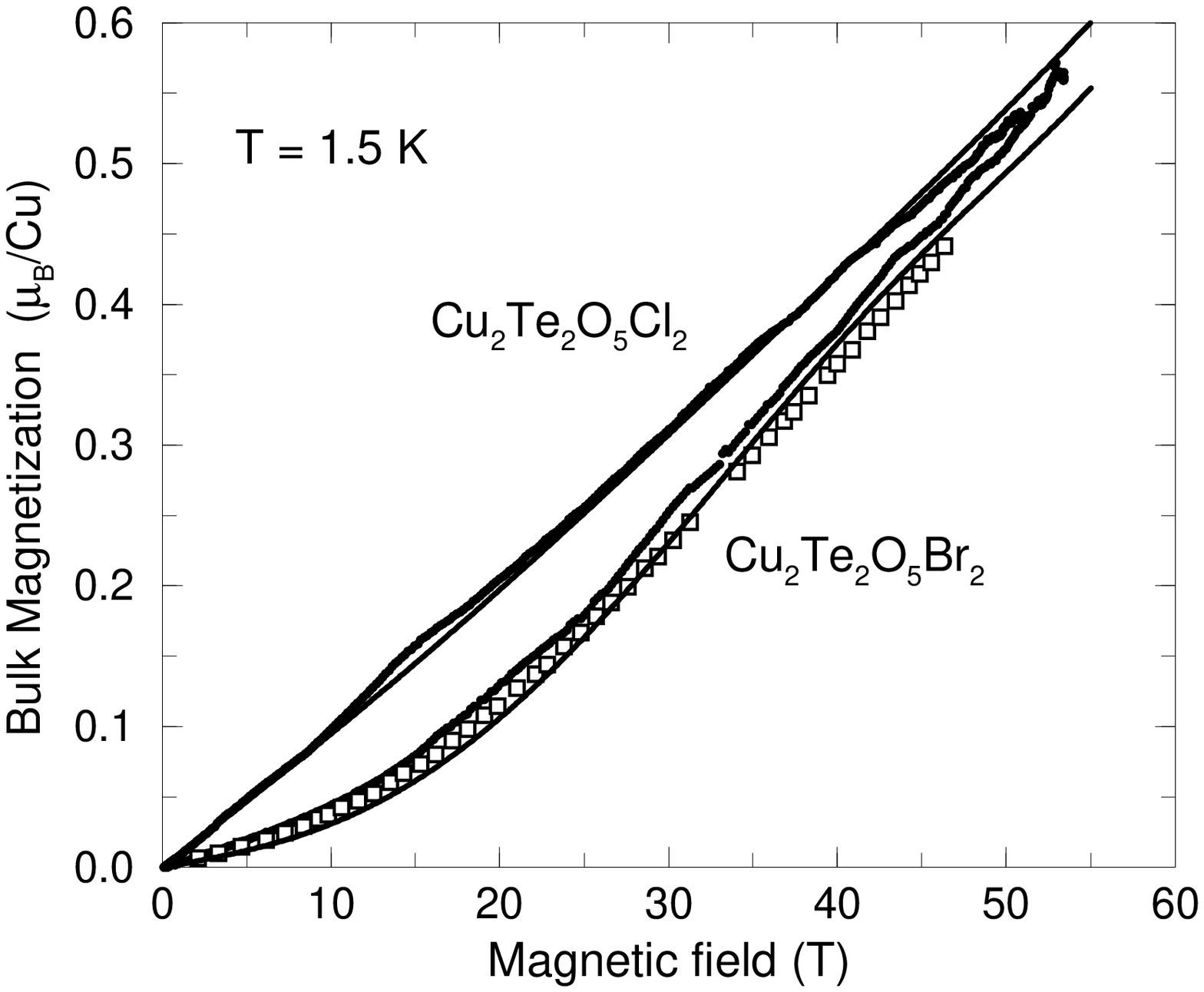}{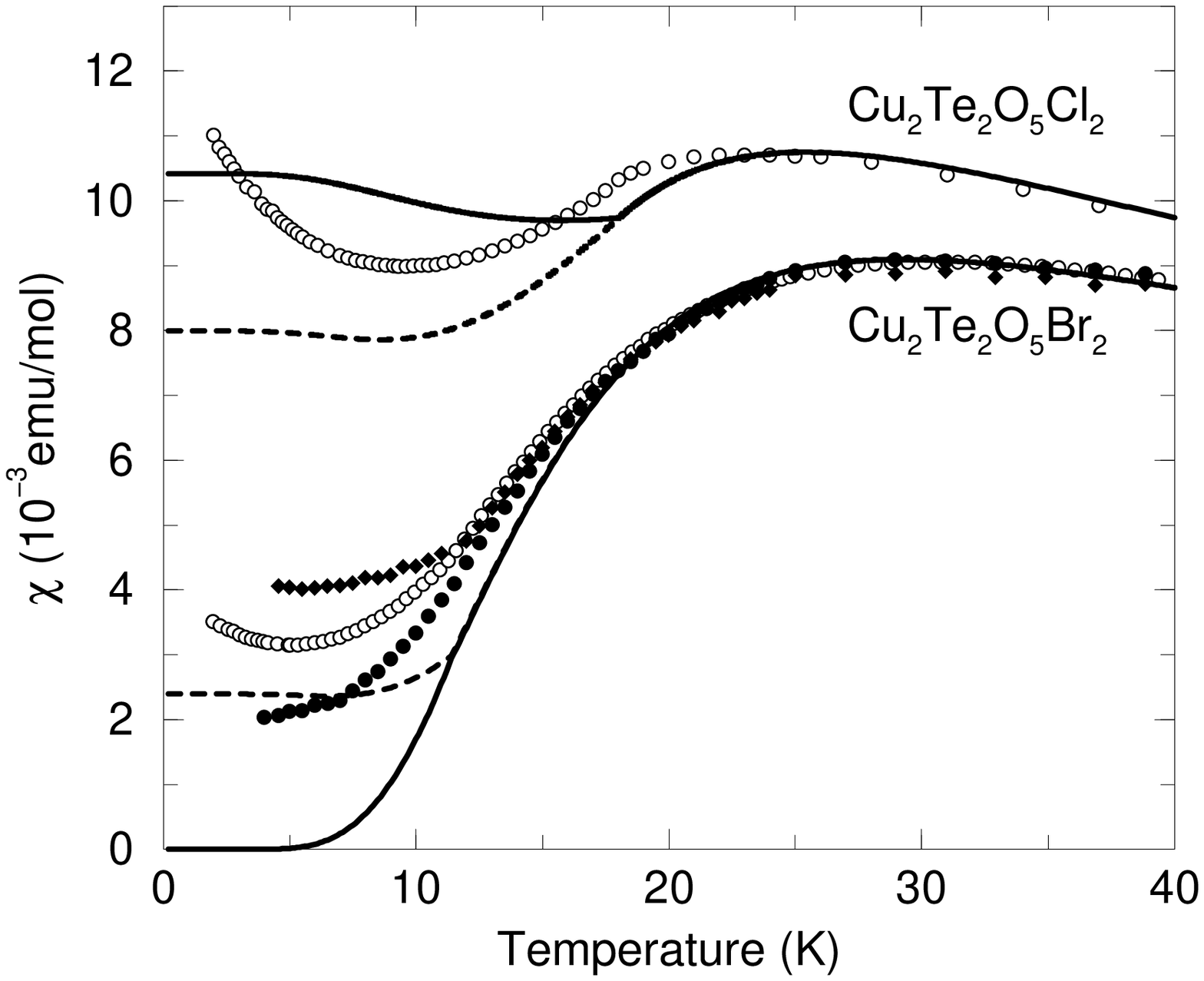}
\caption{High-field magnetization measurements after
ref.~\cite{b.Gros}. The smooth lines are the results of the model
calculations.} \label{f.1}\caption{The bulk susceptibility at low
temperatures. The open circles are the experimental results for
polycrystalline samples \cite{b.PL1,b.Gros}. Single-crystal
results with field parallel and perpendicular to the $c$ axis are
shown (solid symbols) for the case of the Br compound
\cite{b.Gros}. All experimental results have been scaled so that
they approach the theoretical high-temperature value of
$0.7503/T[{\rm K}]$ emu/mol. The solid/dashed lines are the
calculated results when the field is parallel/perpendicular to the
$c$ axis.}\label{f.2} \vspace{10pt}
\twoimages[scale=0.42]{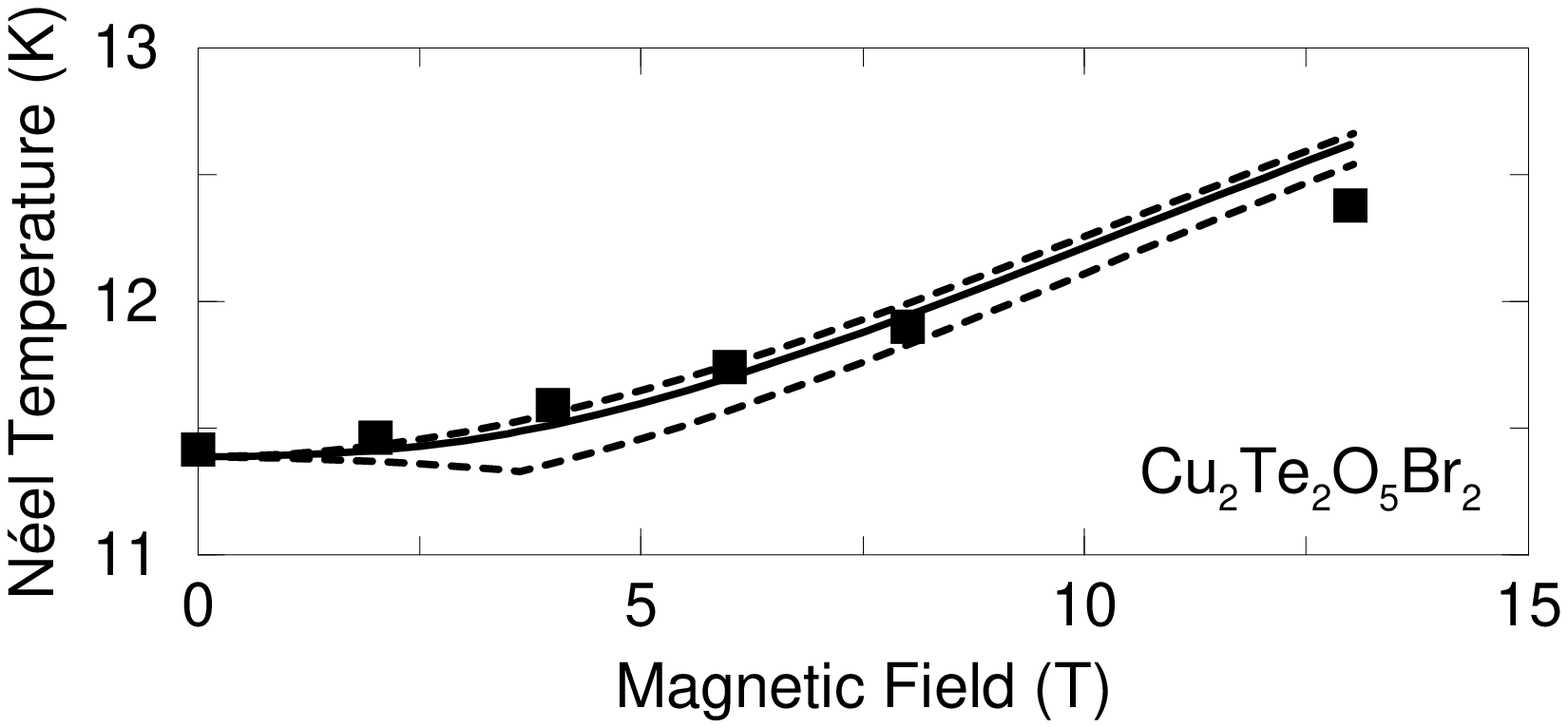}{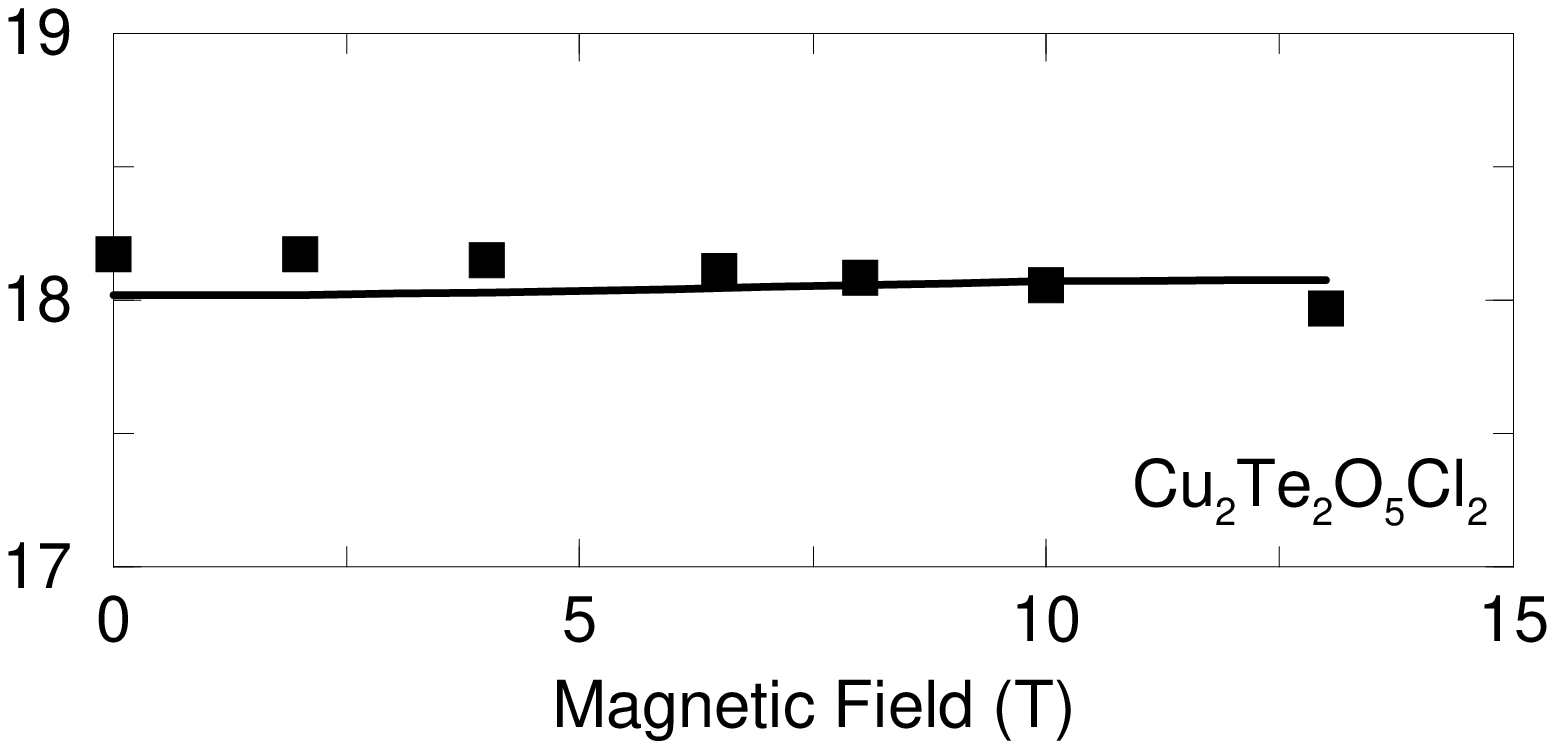}\caption{The
N\'eel temperature as a function of applied field. The dashed
lines (in the Br case) show the calculated \TN\ when the field is
perpendicular (upper one) or parallel (lower one) to the $c$ axis.
The solid lines are the averaged value to be compared with
experimental results \cite{b.PL1,b.Gros} obtained from
polycrystalline samples (the solid squares).} \label{f.3}
\end{figure}

The final model for Cu$_2$Te$_2$O$_5$Br$_2$ includes $J_1^{}$,
$J_2^{}$, $J_c^{}$, and $J_c^{xy}$ defined by the equations
(\ref{e.1})--(\ref{e.3}) plus two additional parameters. $J_f^{}$
is the coupling parameter for the uniform magnetization component,
\begin{equation}
\label{e.9} {\cal H}_Z^{}=-\big(J_f^{}{\bf M}_f^{}+2\mu_B^{}{\bf
H}\big)\cdot\big({\bf S}_1^{}+{\bf S}_2^{}+{\bf S}_3^{}+{\bf
S}_4^{}\big)+2J_f^{}M_f^2,\quad {\bf
M}_f^{}=\frac{1}{4}\big\langle{\bf S}_1^{}+{\bf S}_2^{}+{\bf
S}_3^{}+{\bf S}_4^{}\big\rangle
\end{equation}
of importance in the presence of an external field {\bf H}. The
$S_4^{}$ symmetry of the tetrahedra allows different kinds of
anisotropic interactions. Based on the analysis of the
Raman-scattering experiments we shall concentrate on the following
DM anisotropy term:
\begin{equation}
\label{e.10} {\cal H}_A^{}=D_c^{}\left[({\bf S}_1^{}-{\bf
S}_2^{})\times({\bf S}_3^{}-{\bf S}_4^{})\right]_c^{}
\end{equation}
proportional to the $c$ component of the cross product. The values
of the five (six) parameters in the case of
Cu$_2$Te$_2$O$_5$Br$_2$ are determined to be
($J_c^{xy}<2.17J_1^{}\Rightarrow M_{xy}^{}=0$)
\begin{equation}
\label{e.11}
 J_1^{}=47.5~\hbox{K},\qquad J_2^{}=0.7J_1^{},\qquad
D_c^{}=0.06J_1^{},\qquad J_c^{}=0.856J_1^{},\qquad J_f^{}=0.6
J_1^{}
\end{equation}
The strongest coupling in the system $J_1^{}$ is derived from the
high-temperature susceptibility measurements assuming fixed values
for the remaining ones. The value of $J_f^{}$ is mainly determined
by the high-field magnetization measurements in the ordered phase,
see fig.~\ref{f.1}, but still has a large uncertainty of about
20\%. Due to the $D_c^{}$ anisotropy the bulk susceptibility
becomes anisotropic. The anisotropy is minute in the paramagnetic,
but substantial in the ordered phase. Independent of its sign
$D_c^{}$ implies that the ordered staggered moment $M_z^{}$ in
eq.~(\ref{e.2}) is directed along the $c$ axis. The bulk
susceptibility along this direction is predicted to vanish
exponentially at low temperatures whereas the perpendicular
component stays approximately constant below \TN. This is in
qualitative agreement with experiments, as shown in
fig.~\ref{f.2}. The experimental data are unfortunately somewhat
disturbed by other contributions, as reflected in, for instance, a
sample dependence of the results. The Zeeman energy gained in the
perpendicular case induces a spin-flop transition, when the field
is parallel to the $c$ axis, to a phase where the staggered
moments become perpendicular to the field. The model predicts the
spin-flop field to be 3.7 T, nearly independent of the temperature
below \TN. This spin-flop transition has not yet been observed,
but the anisotropy measurements in fig.~\ref{f.2} were made at a
field of 5 T indicating that the spin-flop field is larger than 5
T. $J_c^{}$ has no influence on the bulk susceptibility, above
\TN, and is adjusted so that the model predicts $\TN=11.4$ K at
zero field. Figure \ref{f.3} shows the transition temperature for
a polycrystalline sample as a function of field. It is remarkable
that \TN\ increases with the field.

\begin{figure}
\twoimages[scale=0.42]{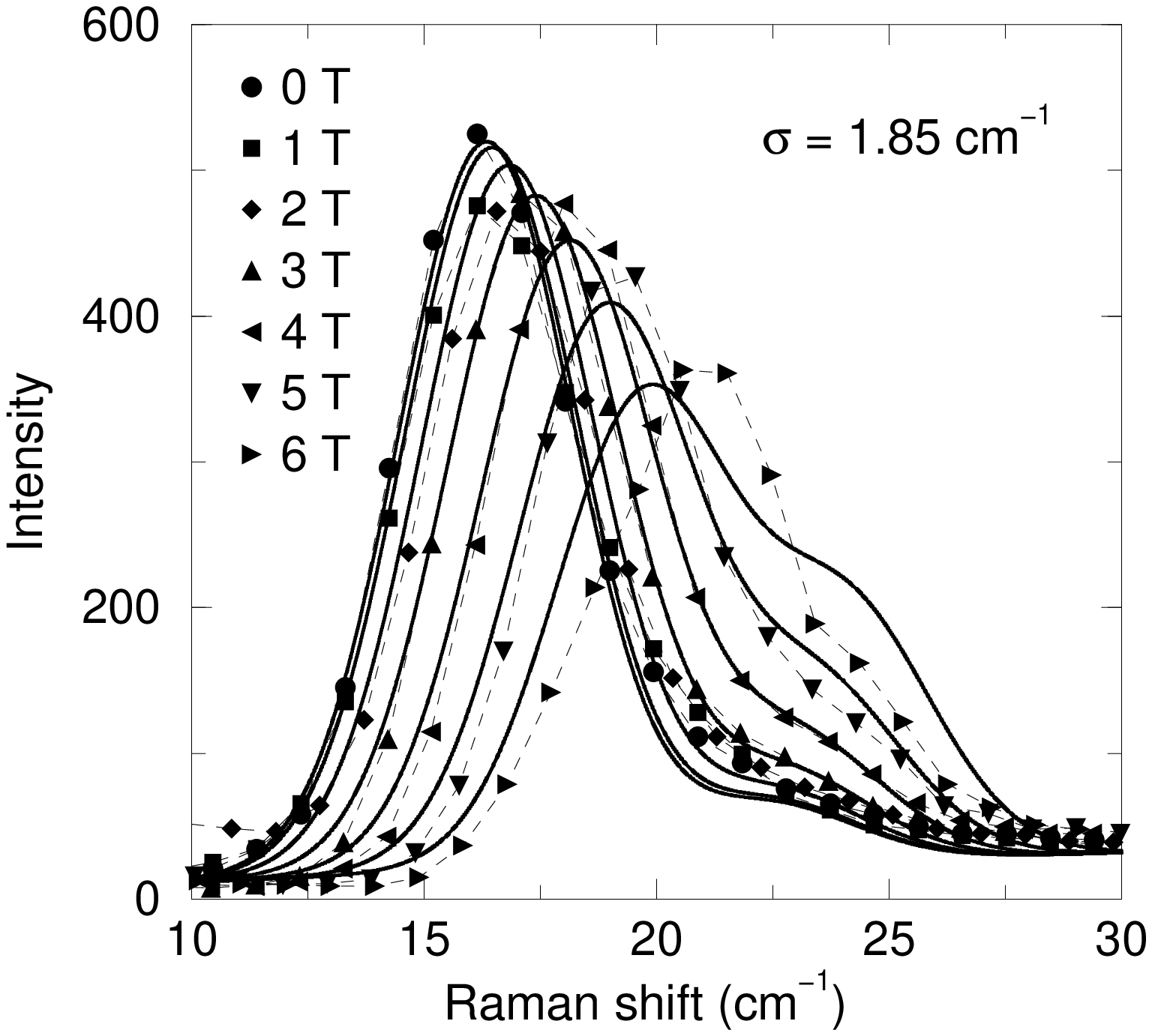}{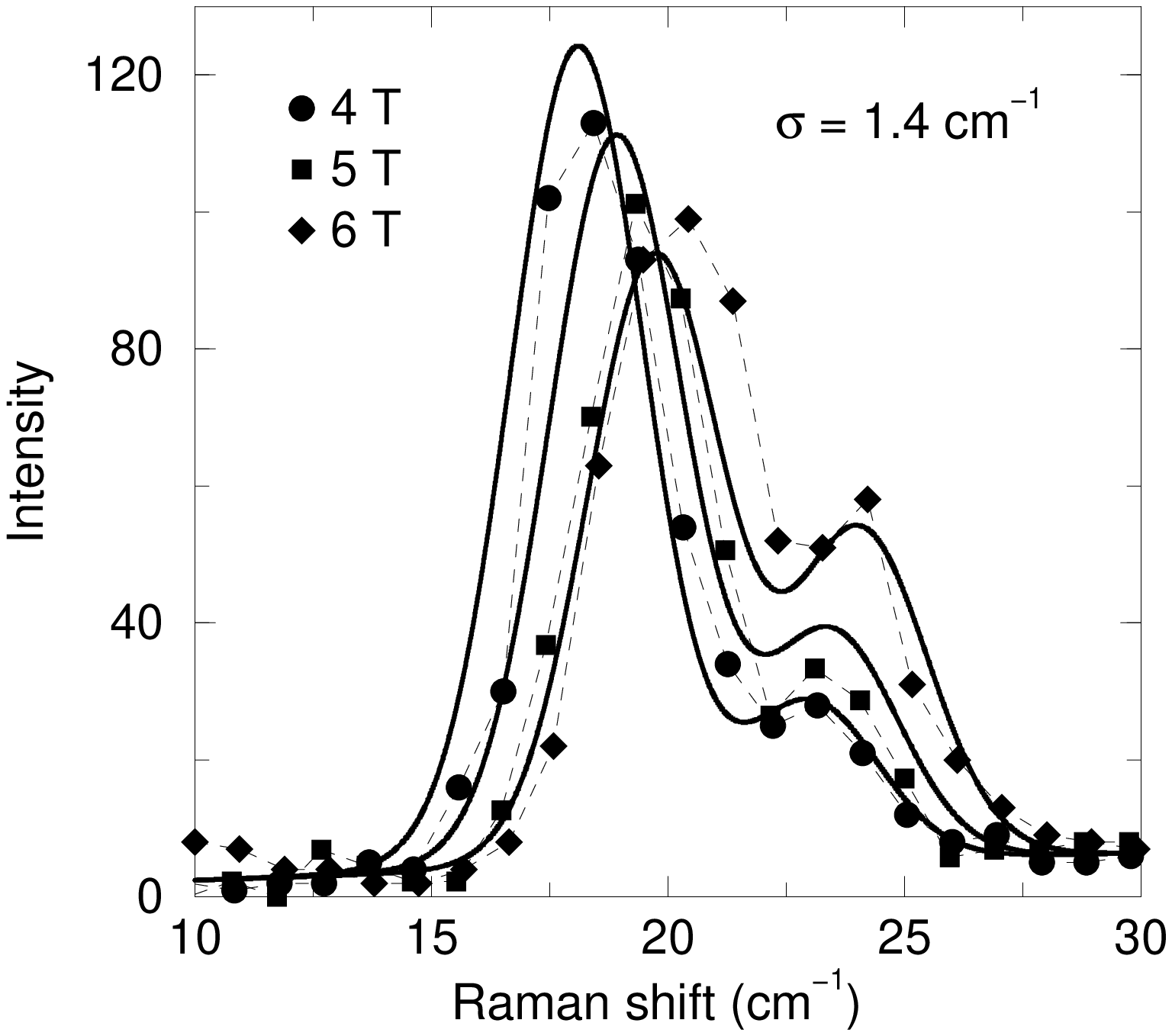} \caption{Raman
spectra of Cu$_2$Te$_2$O$_5$Br$_2$ at 2.1 K in a magnetic field
applied perpendicular to the $c$ axis. The experimental results
(symbols connected by dashed lines) \cite{b.PL1,b.PL2,b.Gros} have
been obtained with two different resolutions and the theoretical
intensities (solid lines) have been folded with Gaussians with the
specified values of $\sigma$. The calculated Raman shifts have all
been subtracted by 1.6 cm$^{-1}$.} \label{f.4}
\end{figure}

The Raman spectra of the Cu$_2$Te$_2$O$_5$(Br$_{1-x}$Cl$_x)_2$
compounds have been studied in detail \cite{b.PL1,b.PL2,b.Gros}.
The longitudinal magnetic excitation at zero wave vector is
observed in the Br system at 2.1 K, when the incoming and
scattered light are polarized along the $c$ axis. This part of the
results is shown in fig.~\ref{f.4}. The high-resolution data
revealed an additional peak at slightly higher energy than the
longitudinal excitation, appearing when a field is applied
perpendicular to the $c$ axis. Gros \etal\ \cite{b.Gros}
interpreted the extra peak to be the $|s_2^{}\rangle$ singlet
level, which is becoming visible due to a DM component in the
cross section combined with the level mixing induced by the field.
The problem with this explanation is that the field-induced mixing
is weak, and a DM cross section as large as the normal part would
only produce a peak of the size of a hundredth of the main one.
The alternative explanation is that it is the Hamiltonian itself,
which is responsible for the mixing of the levels. The only
possible term producing such an effect is the $D_c^{}$ coupling
defined by eq.~(\ref{e.10}). $D_c^{}$ gives rise to a coupling
between the longitudinal mode and the $|s_2^{}\rangle$ level even
at zero field. The application of a field perpendicular to the $c$
axis diminishes the energy difference between the two levels and
the hybridization is strongly increased. The two peaks would
attain equal intensities at a field slightly larger than the
experimental maximum field. Figure \ref{f.4} compares the
intensities predicted by the model, eq.~(\ref{e.8}) with
$\alpha=0$, with the experimental ones. In these calculations
$D_c^{}$ and $J_2^{}$, plus one intensity scale parameter in each
of the two cases, have been utilized as adjustable parameters. The
experimental Raman shifts are subjected to an absolute uncertainty
of the order of 4 cm$^{-1}$ and a relative one of the order of 1
cm$^{-1}$. The relative uncertainty has been reduced by performing
a translation of each spectrum so that the phonon frequencies in
each case coincide with those observed in the sample at zero
field. Because of the large absolute uncertainty we have tried to
mimic the relative behavior rather than to get the right values
for the absolute frequencies. The model thus leads to a good
account of both the frequencies and the intensities if shifting
all the calculated peaks rigidly by $-1.6$ cm$^{-1}$.

\begin{figure}
\onefigure[scale=0.54]{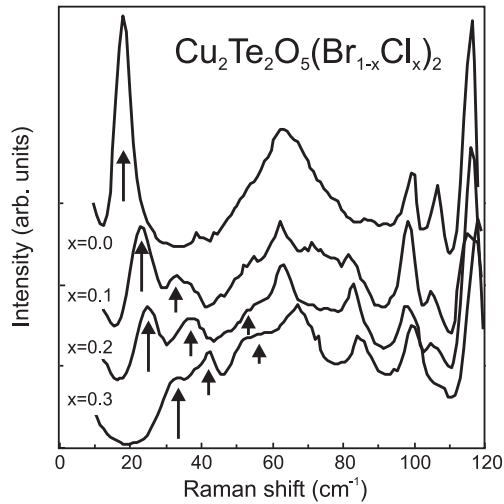} \caption{Raman spectra
obtained for small concentration $x$ of Cl ions at zero field and
4K, after ref.\ \cite{b.PL3}. The arrows indicate the positions of
magnetic scattering peaks. The intensity of the longitudinal
magnetic excitation rapidly decreases with $x$ and new peaks
develop at higher frequencies.} \label{f.5}
\end{figure}

The model we shall present now for the Cu$_2$Te$_2$O$_5$Cl$_2$
system is more speculative. In comparison with the Br compound the
$a$ and $c$ lattice parameters in the Cl system are reduced by 3
and 1\%, respectively, whereas the intra-tetrahedral distances
between the Cu atoms are increased by about 1\%, see ref.\
\cite{b.John}. In correspondence to these changes $J_1^{}$ is
reduced in the Cl compound whereas the N\'eel temperature is
increased. A first guess of a model for the Cl compound would be
the same one as discussed above with some appropriate adjustments
of the coupling parameters. The model based on this assumption
predicts a strong Raman-scattering peak at about 40 cm$^{-1}$, two
to three times as intense as in the Br system if assuming the
scattering coefficients $a_{lm}^{}$ to be the same. However, the
Raman spectra clearly show a quick and systematic reduction of the
magnetic scattering with the concentration $x$ of the Cl ions, see
fig.\ \ref{f.5}. One possible explanation would be that it is the
Raman scattering circumstances, not the model itself, which is
drastically changed. This could be achieved by introducing a
non-zero $\alpha$ in eq.~(\ref{e.7}), which $\alpha$ should then
approach $J_c^{}/2J_1^{}$ in the pure Cl system. We have analyzed
this possibility, but found it to be both implausible and not
capable of reproducing the observed non-monotonic changes of the
Raman spectra with $x$.

The analysis of the magnetization measurements shown in
fig.~\ref{f.1} indicates that $r=J_2^{}/J_1^{}$ is close to 1 in
the Cl compound, so although $J_1^{}$ is reduced $J_2^{}$
increases with the Cl concentration. When $r\approx1$ then
$J_c^{}$ has to be about $2J_1^{}$ in order to produce a
transition to the $M_z^{}$ ordered phase at $\TN=18.2$ K. Another
possibility is that in both systems $J_c^{}\approx J_1^{}$ whereas
$J_c^{xy}\approx1.8J_1^{}$, then the increase of $r$ from about
0.7 in the Br to about 1 in the Cl compound is a sufficient cause
for a change of the order parameter from $M_z^{}$ to $M_{xy}^{}$.
Based on this possibility, the model of Cu$_2$Te$_2$O$_5$Cl$_2$ is
proposed to be ($J_c^{}<1.95J_1^{}\Rightarrow M_z^{}=0$):
\begin{equation}
\label{e.12}
 J_1^{}=40.7~\hbox{K},\qquad J_2^{}=J_1^{},\qquad
D_c^{}=0.06J_1^{},\qquad J_c^{xy}=1.8J_1^{},\qquad J_f^{}=0.6
J_1^{}
\end{equation}
This model produces the magnetization and susceptibility results
displayed in the figs.~\ref{f.1} and \ref{f.2}, and \TN\ is found
to be nearly independent of an external field, as shown in
fig.~\ref{f.3}. Most importantly, the assumption of a $M_{xy}^{}$
instead of a $M_z^{}$ ordering implies a reduction of the
calculated magnetic Raman-scattering intensities by nearly a
factor of 10. Furthermore, instead of one longitudinal mode the
present model predicts two main peaks (at 43 and 75 cm$^{-1}$),
plus a third one (at 57 cm$^{-1}$) observable if the polarization
vectors of the light have a component perpendicular to the $c$
axis. This is, at least qualitatively, in much better agreement
with the Raman experiments \cite{b.PL2,b.PL3} than that predicted
by a $M_z^{}$ ordered phase.

\section{Conclusion}

The analysis by Gros \etal~\cite{b.Gros} of the tetrahedral
spin-$\frac{1}{2}$ cluster system Cu$_2$Te$_2$O$_5$Br$_2$ has been
extended, and the Raman-scattering results have been explained
quantitatively by introducing a Dzyaloshinskii-Moriya anisotropy
term in the Hamiltonian, which term also accounts (partly) for the
anisotropy of the system below \TN. The ordered phase in this
compound is the one, where $\langle{\bf S}_1^{}+{\bf
S}_2^{}\rangle$ is parallel and $\langle{\bf S}_3^{}+{\bf
S}_4^{}\rangle$ antiparallel with the $c$ axis. A spin-flop
transition is predicted to occur when applying a field along the
$c$ axis. It is worth pointing out that the magnetic Raman
spectrum (leaving out the two-magnon scattering part) depends on
the resulting mean field, but not on the relative orientation of
the magnetic moments on neighboring tetrahedra. A determination of
this would require a neutron-diffraction investigation. The
present analysis of the Raman spectra indicates that the ordering
in Cu$_2$Te$_2$O$_5$Cl$_2$ compound is of the $J_2^{}$-type, where
$\langle{\bf S}_1^{}-{\bf S}_2^{}\rangle$ and $\langle{\bf
S}_3^{}-{\bf S}_4^{}\rangle$ are perpendicular to each other and
to the $c$ axis. The evidences are indirect and the
$J_2^{}$-structure needs to be verified by diffraction
experiments.

The mean-field/random-phase approximation is a relevant starting
point, whenever the interactions are strong enough to induce
magnetic ordering. This is also true close to quantum criticality
(Pr-metal may serve as a good example for that, \cite{b.Jens}).
The non-mean-field behavior of the present frustrated system
appears to be accounted for by the treatment of each tetrahedral
Cu-spin cluster as a local dimerized unity. The analysis shows
that the specification of the properties of the local spin
clusters requires at least three parameters $J_1^{}$, $J_2^{}$,
and $D_c^{}$. The values of the three MF parameters $J_f^{}$,
$J_c^{}$, and $J_c^{xy}$, which correspond to three different
combinations of the exchange interactions between neighboring
tetrahedra, indicate that both the interactions perpendicular and
parallel to the $c$-axis chains of tetrahedra are important and
therefore that the system is three- rather than one-dimensional.
The good coincidence between the predictions of the MF/RPA theory
and the experimental results is a further argument for a
three-dimensionality of the magnetic interactions in the two
compounds.

\acknowledgments
We thank K.Y. Choi for important discussions and DFG SPP 1073 for
financial support.

\end{document}